\documentclass[twocolumn,twoside,slac_two]{revtex4}
\usepackage{graphicx}
\usepackage{fancyhdr}
\usepackage{hyperref}
\hypersetup{
    colorlinks=true,
    urlcolor=magenta,
}
\begin{document}

\title{Results on Solar Physics from AMS-02}

%

\author{S. Della Torre and AMS-02 Collaboration}
\affiliation{INFN Milano-Bicocca, Piazza della Scienza 3, 20125 Milano, Italy}
%

\begin{abstract}
AMS-02 is a wide acceptance high-energy physics experiment installed on the International Space Station in May 2011 and operating continuously since then. 
Using the largest number of detected particles in space of any space-borne experiment, it performs precision measurements of  galactic cosmic rays fluxes.
Detailed time variation studies of Protons, Heliums, Electron and Positron fluxes were presented. 
The low-rigidity range exhibits a decreasing general trend strongly related to the increase of solar activity, as well local decreases associated with strong solar events.
\end{abstract}

\maketitle

\thispagestyle{fancy}

\begin{figure}[bt]
\includegraphics[width=0.48\textwidth]{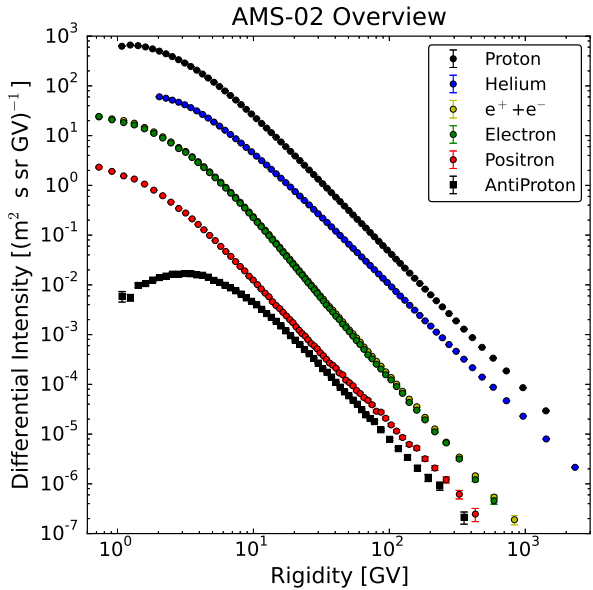}
\caption{AMS-02 measured spectrum for proton (black points)~\cite{2015PhRvL.114q1103A},
helium (blue points)~\cite{2015PhRvL.115u1101A}, electron (green points)~\cite{2014Aguilar}, 
Positron (red points)~\cite{2014Aguilar}, e$^-$+e$^+$ (yellow points)~\cite{2014PhRvL.113v1102A} 
and antiproton (black squares)~\cite{2016PhRvL.117i1103A}.
}
\label{Fig::AMS02_publicResult}
\end{figure}

The Alpha Magnetic Spectrometer (AMS-02) was installed on the International Space Station
(ISS)
to measure cosmic rays with unprecedented accuracy. 
The long duration of mission, planned to conclude in 2024 with ISS mission, allows to cover almost a complete solar cycle
from the ascending phase of this solar cycle 24, through its maximum, and the descending phase into the next solar minimum. 
Thanks to its large geometrical acceptance, 0.45 m$^2$ sr, over 65 billion
cosmic ray events have been recorded in the first 48 months
of AMS operations~\cite{2016PhRvL.117i1103A}.
A collection of all AMS-02 available spectra is shown in Fig.~\ref{Fig::AMS02_publicResult} where black point are proton spectrum~\cite{2015PhRvL.114q1103A},
blue points are helium~\cite{2015PhRvL.115u1101A}, green point are electron~\cite{2014Aguilar}, 
red points are positron~\cite{2014Aguilar},yellow points represents e$^-$+e$^+$~\cite{2014PhRvL.113v1102A} (almost overlapped with electron flux) 
and finally black squares are antiproton~\cite{2016PhRvL.117i1103A}

These data will allow deep studies of the Galactic Cosmic Ray
(GCR) fluxes and their time evolution over an entire solar cycle enabling a better understanding of
the so-called \textit{solar modulation} effect. 
In this paper we reviewed the AMS-02 analysis for proton, helium, electron and positron. 
We presented the time variation of GCR flux during first 5 years of data taking with particular attention of range 1--30 GV.
We explore long term effects on GCR flux with different charge and masses. 
Finally, short scale solar events are presented. 

\section{AMS-02 detector}
\begin{figure}[bt]
\includegraphics[width=65mm]{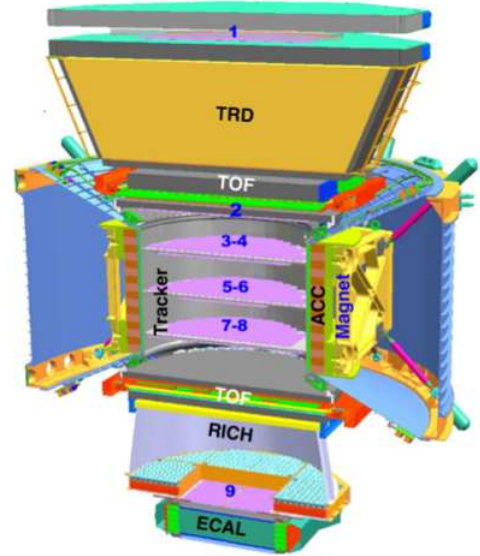}
\caption{Schematic view of AMS-02 detector. Sub-detectors from top to bottom: transition radiation
detector (TRD); time of flight (TOF); tracker (9 layers) and magnet; anti-coincidence counters (ACC); ring
imaging Cerenkov detector (RICH); electromagnetic calorimeter (ECAL)}
\label{Fig::AMS02}
\end{figure}
The AMS-02 is a general purpose
particle physics detector, operating in space since May
2011. It will achieve a unique long duration mission,
aiming at performing antimatter and dark matter searches,
as well as cosmic ray composition and flux measurements in the rigidity range from 1 GV to a few TV,
where rigidity is the momentum divided by the charge.
The experiment is installed on-board the ISS, 
located at an average altitude of about 400 Km from Earth surface, thus over the atmosphere but inside the geomagnetic field.

In order to distinguish between different species
of particles, AMS-02 is made of 5 sub-detectors which are described in detailed in Ref \cite{2013PhRvL.110n1102A}.
From top to bottom (see Fig. \ref{Fig::AMS02}) the detector is composed 
of a transition radiation detector (TRD), a time of flight system (TOF), 
nine planes of precision silicon tracker surrounded by a permanent magnet, 
an array of anti-coincidence counters (ACC), 
a ring imaging \u{C}erenkov detector (RICH), and an electromagnetic
calorimeter (ECAL).

The tracker accurately determines the trajectory and
absolute charge (Z) of cosmic rays by multiple 
measurements of the coordinates and energy loss. It is composed of
192 ladders, each containing double-sided silicon sensors,
readout electronics, and mechanical support~\cite{2010NIMPA.613..207A,2011Tracker}. 
Three planes of aluminum honeycomb with carbon fiber skins are
equipped with ladders on both sides of the plane. These
double planes are numbered 3--8 (see Fig.~\ref{Fig::AMS02}). 
Another three planes are equipped with one layer of silicon ladders. 
As indicated in Fig.~\ref{Fig::AMS02}, plane 1 is located on top of the TRD,
plane 2 is above the magnet, and plane 9 is between the
RICH and the ECAL. Plane 9 covers the ECAL acceptance. 
Planes 2--8 constitute the inner tracker. Coordinate
resolution of each plane is measured to be better than
10 $\mu$m in the bending direction, and the charge resolution
is $\Delta Z \simeq 0.06$ at Z=1. The total lever arm of the tracker
from plane 1 to plane 9 is 3.0 m. Positions of the planes of
the inner tracker are held stable by a special carbon fiber
structure~\cite{2002Tracker}. It is monitored by using 20 IR laser beams
which penetrate through all planes of the inner tracker and
provide micron-level accuracy position measurements.
The positions of planes 1 and 9 are aligned by using cosmic
ray protons such that they are stable to 3 $\mu$m (see, e.g., Figure 2 in Ref.~\cite{2013PhRvL.110n1102A}).
Together with the tracker, the magnet provides a maximum detectable 
rigidity of 2 TV on average, over tracker planes 1--9.

The TRD is designed to use transition radiation to distinguish between $e^\pm$ and protons, and $dE/dx$ 
to independently identify nuclei \citep{2006TRD}. It consists of 5\,248 proportional
tubes of 6 mm diameter with a maximum length of 2 m
arranged side by side in 16-tube modules. The 328 modules
are mounted in 20 layers. Each layer is interleaved with a
20 mm thick fiber fleece radiator with a density
of 0.06 g/cm$^3$. The tubes are filled with a 90:10 Xe:CO$_2$ mixture.
The on-board gas supplies contained, at launch,
49 kg of Xe and 5 kg of CO$_2$ which ensures $\simeq$ 30 years of
steady TRD operations in space. In order to differentiate
between $e^\pm$ and protons, signals from the 20 layers are
combined in a TRD estimator formed from the ratio of the
log-likelihood probability of the $e^\pm$ hypothesis to that of
the proton hypothesis. Positrons and electrons have a TRD
estimator value $\sim 0.5$ and protons $\sim 1$. This allows the
efficient discrimination of the proton background.

Two planes of TOF counters are located above and two
planes below the magnet \citep{2013ToF,2014NIMPA.743...22B}. 
Each plane contains eight or
ten scintillating paddles. Each paddle is equipped with two
or three photomultiplier tubes on each end for efficient
detection of traversing particles. The coincidence of signals 
from all four planes provides a charged particle trigger. 
The TOF charge resolution, obtained from multiple
measurements of the ionization energy loss, is $\Delta Z \simeq 0.05$
at $Z= 1$. The average time resolution of each counter has
been measured to be 160 ps, and the overall velocity
($\beta = v/$c) resolution of the system has been measured to
be 4\% for $\beta \simeq  1$ and $Z = 1$ particles, which also discriminates 
between upward- and downward-going particles.
The timing resolution improves with increasing magnitude
of the charge to a limit of $\Delta t \sim 50$ ps and $\Delta \beta/\beta \sim 1$\% for
$Z > 5$ particles.

The ACC surround the inner tracker inside the
magnet bore \cite{2009ACC}. Their purpose is to detect events with
unwanted particles that enter or leave the inner tracker
volume transversely. The ACC consists of sixteen curved
scintillator panels of 0.8 m length, instrumented with
wavelength-shifting fibers to collect the light. 
Long duration tests of the counters show
they have an efficiency close to 0.999\,99.

The RICH is designed to measure the magnitude of the
charge of cosmic rays and their velocities with a precision
of $\Delta \beta/ \beta \sim 1/1000$ \cite{2010Aguilar}.
It consists of two non-overlapping
dielectric radiators, one in the center with a refractive
index of $n= 1.33$, corresponding to a \u{C}erenkov threshold
of $\beta > 0.75$, surrounded by a radiator with $n = 1.05$, with
a threshold of $\beta > 0.95$. 

The ECAL consists of a multilayer sandwich of 98 lead
foils and $\sim50\,000$ scintillating fibers with an active area of
$648 \times 648$ mm$^2$ and a thickness of 166.5 mm corresponding 
to 17 radiation lengths \cite{2012Rosie}. The calorimeter is composed of nine superlayers, 
each 18.5 mm thick and made of
11 grooved, 1 mm thick lead foils interleaved with
ten layers of 1 mm diameter scintillating fibers (the last
foil of the last superlayer is made of aluminum). In each
superlayer, the fibers run in one direction only. The 3D
imaging capability of the detector is obtained by stacking
alternate superlayers with fibers parallel to the x and y axes
(five and four superlayers, respectively). 
In order to cleanly identify electrons and positrons, an ECAL estimator, based on a
boosted decision tree algorithm, is constructed by
using the 3D shower shape in the ECAL.

There are three main detectors that allow a significant
reduction of the proton background in the identification of
the positron and electron samples. These are the TRD
(above the magnet), the ECAL (below the magnet), and
the tracker. The TRD and the ECAL are separated by the
magnet and the tracker. This ensures that secondary particles 
produced in the TRD and the upper TOF planes are
swept away and do not enter into the ECAL. Events with
large angle scattering are also rejected by a quality cut on
the measurement of the trajectory using the tracker. The
matching of the ECAL energy and the momentum measured 
with the tracker greatly improves the proton rejection. 
The proton rejection power of the TRD estimator at
90\% $e^\pm$ efficiency measured on orbit is $10^3-10^4$. 
The proton rejection power of the ECAL estimator
when combined with the energy-momentum matching
requirement $E/p > 0.75$ reaches $\sim10\,000$. The
performance of both the TRD and ECAL estimators are
derived from data taken on the ISS. Note that the proton
rejection power can be readily improved by tightening the
selection criteria with reduced $e^\pm$ efficiency.

\section{Flux Measurement and Analysis}
\begin{figure}
\includegraphics[width=0.48\textwidth]{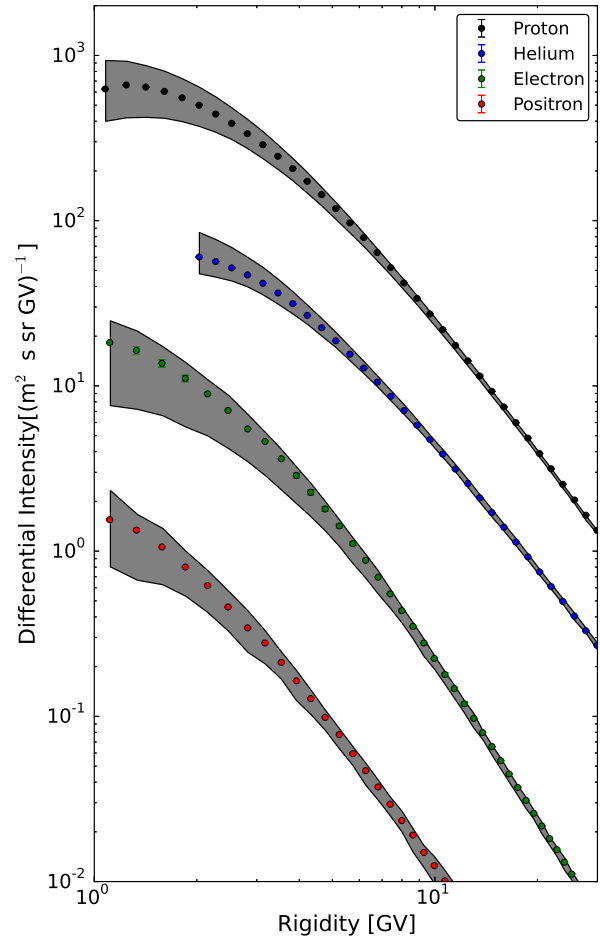}
\caption{AMS-02 measured spectrum for proton (black point)~\cite{2015PhRvL.114q1103A},
helium (blue) point~\cite{2015PhRvL.115u1101A}, electron (green point)~\cite{2014Aguilar} and 
Positron (red point)~\cite{2014Aguilar}. The gray area represent the total flux variation during the first 5 year of AMS-02 operations.}
\label{Fig::ProtRigAll}
\end{figure}
In this work the AMS-02 data collected during first 5 years of AMS-02 operations are analyzed.
The isotropic flux of GCRs in each Rigidity bin $R$ (expressed in GV), of width $\Delta R$, is given
by~\cite{2015PhRvL.114q1103A}:
\begin{equation}
 \Phi(R)=\frac{N(R)}{A_{eff}\cdot T(R) \cdot \Delta R}
\end{equation}
where $N(R)$ is the number of identified particles with
rigidity between R and $R + \Delta R$, $A _{eff}$ is the effective acceptance,
$T(R)$ is the exposure time. 
The effective data collection time i.e. exposure time, is measured by
requiring that AMS-02 is in the nominal data taking status, the detector pointing direction is within
40$^\circ$ of the Earth zenith axis, and the ISS is orbiting outside the South Atlantic Anomaly. 
Each GCR species was analyzed by independent groups with the aims to reduce systematic uncertainties.
The integratedflux for each species can be found in Ref.~\cite{2015PhRvL.114q1103A,2015PhRvL.115u1101A,2014Aguilar,2014PhRvL.113v1102A,2016PhRvL.117i1103A} and are reported in Fig.~\ref{Fig::AMS02_publicResult}.

\subsection{Proton Analysis}
The proton sample is selected by requiring a downward going event with measured $\beta > 0.3$.
The charge measurement must be consistent with a charge one particle, $|Z|=1$, along the full particle trajectory. 
The measured rigidity must be positive and above the International Geomagnetic
Reference Field (IGRF\cite{2010IGRF}) cutoff according to the maximum value in the AMS-02 field of view.
To eliminate further contamination of secondary particles coming from the penumbra region, an
additional 1.2 safe factor was applied to the IGRF cutoff. 
To remove events with large scattering
and to further clean the data sample, additional track fitting requirements on the $\chi^2$ in the bending
coordinate were included. Finally a cut on the combined TOF and tracker mass measurement ($m >
0.5$ GeV/c$^2$) was applied to suppress the small contamination of secondary pions produced in the
upper part of the detector.
Protons are the most abundant species of GCRs, so the residual background from other particles is very low. 
Deuterons are not removed in this analysis: their contamination is less than 2\%
at 1 GV and decreases with increasing rigidity (0.6\% at 20 GV). Contamination from interacting
nuclei ($Z>1$) at the top of AMS-02 (layer 1 or TRD) is 0.5\% at 1 GV and becomes negligible with
increasing rigidity. Positron and electron contamination is less than 0.1\% at all rigidities.
All detector efficiencies have been extensively studied and validated with ISS data. The
trigger efficiency indeed is directly measured from ISS data by the use of a prescaled (1\%) unbiased
trigger sample with no ACC requirement and a coincidence of at least 3 out of the 4 TOF layers.
The number of events was corrected with the rigidity resolution function to account for the
bin-to-bin migration, i.e. unfolding procedure. Two unfolding procedures were used to crosscheck
the result: the so called folded acceptance and the forward unfolding technique whose details are
described in Ref. \cite{1983Unfolding} and Ref. \cite{2007Unfolding} respectively. 
The small difference between the two methods (less
than 0.5\%) is accounted into the systematic errors.
The systematic errors have been extensively studied and a detailed description is given in
Ref. \cite{2015PhRvL.114q1103A}. The order of magnitude of the systematic errors in the rigidity range from 1 GV to 10 GV is summarized here. One source of systematic errors comes from the uncertainties on the
trigger efficiency due to the reduced statistics of the prescaled unbiased sample: this systematic is
very low ($< 0.4$\%) below 10 GV. Another source of systematic errors comes from the data to MC
corrections that have to be applied to the acceptance: these corrections dominate our systematics
at low rigidities being about 5\% at 1 GV and decreasing below 2\% at 10 GV. In addition, the
acceptance suffers from the uncertainties on the model estimation of the inelastic cross sections.
The total systematic on this latter was estimated to be about 1\% at 1 GV and decreases to 0.6\%
at 10 GV. Finally the uncertainty on the rigidity scale resolution function, which is coming from a
residual tracker misalignment and from the magnetic field map measurement, was found to be less
than 1\% below 10 GV.
To compute fluxes each bartel rotation, additional analysis were performed to verify the detector stability
versus time. 
The daily trigger and both inner and full span tracker efficiencies in the
rigidity range from 1 GV to 10 GV, was showed in Figure 2 of Ref.~\cite{2015ICRC_Consolandi}.
The trigger efficiency remained stable over time during all the period of operations. The
inner and full span tracker efficiencies increased on July 21$^{st}$, 2011 due to the improvement of the
tracker calibration and had a drop on December 1$^{st}$, 2011 due to the loss of 3\% tracker readout
channels in the non-bending coordinates. 
The detector acceptance
and all the efficiencies were calculated for each month to obtain the monthly fluxes. The monthly
proton flux above 45 GV shows no observable effects related to the solar modulation and remained
stable for this measurement period as reported in Ref. \cite{2015PhRvL.114q1103A}.
In Fig.~\ref{Fig::ProtRigAll} the integrated flux presented in 
Ref.~\cite{2015PhRvL.114q1103A} (black round) is showed with the total flux variation of proton for the considered period as the gray band.

\begin{figure}[tb]
\includegraphics[width=0.48\textwidth]{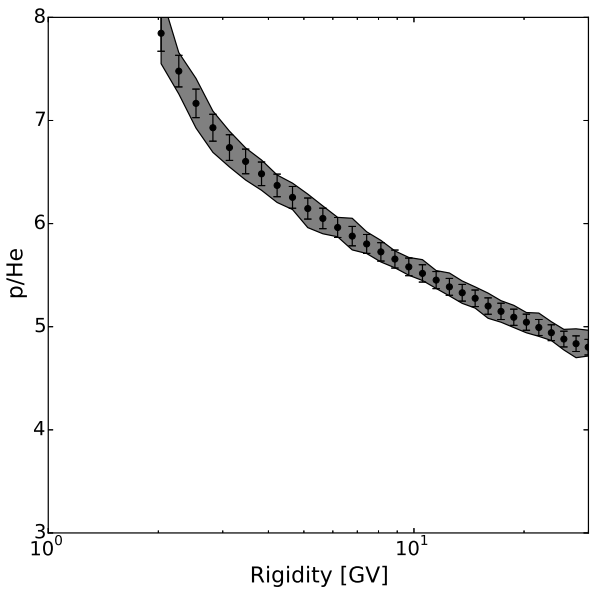}
\caption{Proton over helium flux ratio. 
Black round points are the ratio between 
integrated fluxes presented in Refs.\cite{2015PhRvL.114q1103A,2015PhRvL.115u1101A}.
The gray area is the overall variation along the five year of AMS operation.}
\label{Fig::ProtRatiopHeAll}
\end{figure}

\subsection{Helium analysis}
Helium sample was selected requiring to be downward going and to have a
reconstructed track in the inner tracker with charge compatible with
Z=2. In order to have the best resolution at the
highest rigidities, further selections are made by requiring
the track to pass through L1 and L9 and to satisfy additional
track fitting quality criteria such as a $\chi^2$ cut in the bending
coordinate. To remove the helium candidates
which interacted  within the  detector, were used additional requirements on the  charge  as
measured by each of L1, the upper TOF, the lower TOF,
and L9. To select only primary cosmic rays, the
measured rigidity is required to be greater than a factor of
1.2 times the maximum geomagnetic cutoff within the
AMS field of view. 
Because of the multiple independent measurements of
the charge, the selected sample contains only a small
contamination of particles which had Z$\neq$2 at the top of
the AMS. Comparing the proton and helium charge
distributions in the inner tracker, the proton contamination
of the helium sample was measured to be less than $10^{-4}$
over the entire rigidity range. The sample also contains
helium from other nuclei which interact at the top of the
AMS (for example, in L1), this contribution is below $10^{-3}$
for the entire rigidity range. 
The background contributions are subtracted from the flux and the uncertainties
are accounted for in the systematic errors.
The trigger efficiency was measured
to range from 95\% to 99.5\%, where the inefficiency is
mostly due to secondary $\delta$ rays produced by He in the
tracker materials and which then entered the ACC. 
The bin-to-bin migration of events was corrected using the
two unfolding procedures used in the proton analysis.
Extensive studies, reported in~\cite{2015PhRvL.115u1101A}, were made for the systematic errors.
These errors include the uncertainties in the trigger efficiency, 
the geomagnetic cutoff factor, the acceptance taking
into account the event selection and reconstruction and also
accounting for helium interactions in the detector, the
unfolding, the rigidity resolution function, the absolute
rigidity scale, and the negligible background contamination
discussed above. 
In Fig.~\ref{Fig::ProtRigAll} the integrated flux presented 
in Ref.~\cite{2015PhRvL.115u1101A} (blue round) is showed with 
the total flux variation for the considered period as the gray band.

\subsection{Lepton analys}
The measurement of the separate fluxes of electrons and
positrons is needed for a deeper understanding of the positron fraction
measurement reported in Ref.~\cite{2014PhRvL.113l1101A}. 
The isotropic flux of cosmic rays electrons
and positrons in each energy bin $E$, of width $\Delta E$, is given
by \cite{2014Aguilar}:
\begin{equation}
 \Phi(E)=\frac{N_e(E)}{A_{eff}\cdot T(E) \cdot \Delta E}
\end{equation}
where $N_e(E)$ is the number of electrons or positrons with
energy between E and $E + \delta E$, $A _{eff}$ is the effective acceptance,
$T(E)$ is the exposure time. The effective acceptance
$A _{eff}$ is the product of the detector geometric acceptance
($\sim$ 500 cm$^2$ sr) and the selection efficiency, estimated with
simulated events and validated with a pure sample of electron events identified in the data. The trigger efficiency is
100\% above few GeV, and it is estimated using minimum
bias triggered events. The exposure time is evaluated as
a function of energy and it takes into account the lifetime
of the experiment which depends on its orbit location and
on the geomagnetic cutoff. To identify downward-going
particles of charge one, cuts are applied on the velocity
measured by the TOF and on the charge reconstructed by
the tracker, the upper TOF planes, and the TRD. 
To reject positrons and electrons produced by the interaction of
primary cosmic rays with the atmosphere, the minimum
energy within the bin is required to exceed 1.2 times the
geomagnetic cutoff. Over a sample of well reconstructed
particles with one shower in the ECAL and one track in the
TRD and in the tracker, the identification of signal events
is performed applying an additional cut on E/p, followed
by a fixed cut in the ECAL estimator to further reduce
the proton background. The number of signal and 
background event is estimated for each energy bin performing
a template fit procedure. 
The energy resolution of the ECAL is below 2\% at energies higher than 80 GeV~\cite{2013Adloff} and the absolute energy scale
is verified by using minimum ionising particles and the ratio between the energy, measured by the ECAL, and the
momentum, measured by the tracker. These results are
compared with the Test Beam values where the beam en-
ergy is known to high precision. Between 10 and 290 GeV
(Test Beam energies), the uncertainty on the absolute scale
is $\sim 2\%$, while it is 4\% up to 700 GeV. The statistical error
dominates above 50 GeV in the measurement of positrons,
while the systematic and statistical errors are comparable
for electrons, above 200 GeV (see table 1 in Ref.\cite{2014Aguilar}). 
In Fig.~\ref{Fig::ProtRigAll} the integrated flux presented 
in Ref.~\cite{2014Aguilar} for electron (green points) and positron (red points) are showed with 
the total flux variation for the considered period as the gray band.

\begin{figure}
\includegraphics[width=0.48\textwidth]{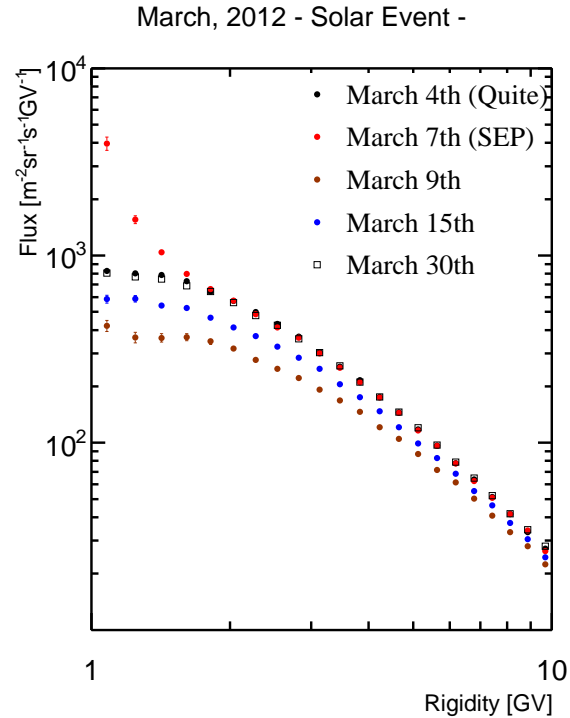}
\caption{Proton flux measured in daily binning during the Solar event of March 7$^{\rm th}$, 2012. As reference March 4$^{\rm th}$ is showed as unperturbed (quite) spectrum.
At March 7$^{\rm th}$, the increase in the
lowest rigidity bin indicates the arrival of SEP. The Minimum of the flux is reached at March 9$^{\rm th}$, 
then the slowly increase to unperturbed flux that last till  March 30$^{\rm th}$}.
\label{Fig::SEPFlux}
\end{figure}
\begin{figure*}
\includegraphics[width=0.9\textwidth]{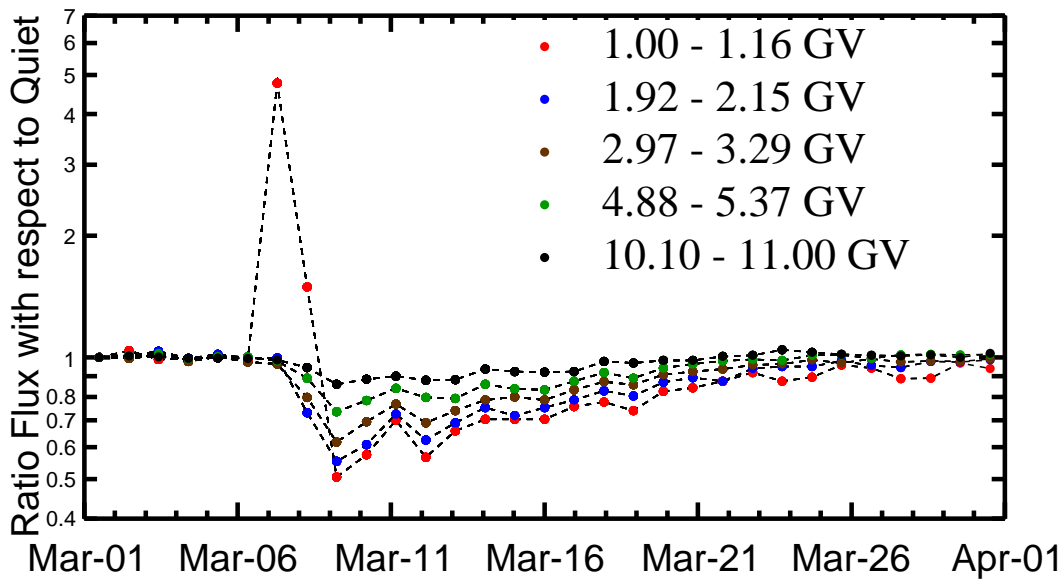}
\caption{ Time
evolution of the daily proton flux observed by AMS-02 during the March 2012 month for 1, 2, 3, 5 and 10 GV. (see text for description)}
\label{Fig::SEPRAtio}
\end{figure*}

\section{Helium over Proton Ratio}
The overall variation of protons and helium is shown in Fig.~\ref{Fig::ProtRigAll}. 
During the first five year of operation the flux decreased down to 50\% with respect the first month. 
AMS-02 operates along the solar maximum of solar cycle 24 and observed the maximum of solar activity in 2014, 
i.e. the minimum of the flux. After that, GCR flux of both proton and helium rise again.
In Fig.~\ref{Fig::ProtRatiopHeAll} we report the ratio of integrated proton and helium flux as presented in 
Ref.~\cite{2015PhRvL.115u1101A} compared with the overall variation along the five year of AMS operation represente as gray bands. 
It is possible to note as the overall variation is compatible with the experimental uncertainties of integrated flux. 
Thus, the helium over proton flux ratio evaluated for same rigidity bin is time independent. 
It is important to note that helium has double charge and four time the mass of proton. 
The evidence of time independence of p/He suggest that the propagation process scale with rigidity.


\section{Electron over Proton Ratio}
The overall variation of electrons and positrons fluxes are shown in Fig.~\ref{Fig::ProtRigAll}. 
Although, since 2011, both electron and positron flux decreased down to 50\% with respect the first month, 
positrons and electrons fluxes shown different time evolution after the peak of solar activity. 
The minimum of electron flux was delayed of more than 12 month with respect positron. 
As a consequence, the positron fraction  -- that, as reported in Ref.~\cite{2014PhRvL.113l1101A}, 
was stable in time during the first 2 year of AMS-02 operation -- from second half of 2013 start to slowly 
change toward results compatible with the one measured by AMS-01~\cite{2002PhR...366..331A} in 1998.
The same is observed comparing protons and electron fluxes, while protons and positrons seems to follow the same time evolution.
This indicates a charge sign dependent effect, that is negligible before 2013 and becomes important moving towards solar minimum.

\section{Solar Energetic Particles}

In addition to this overall modulation effect, shorter timescale fluctuations are observed. 
These  are related to solar impulsive events. Sharp increase of AMS low energy flux are associated 
to Solar Energetic Particles (SEP) penetrating the Earth magnetic field. 
A decrease in AMS low energy flux, that last for few days, is usually labeled as Forbush Decrease (FD).
A partial list of events that were observed by AMS-02 detector can be found in Ref.~\cite{2015ICRC_Bindi}.
For this analysis the computed geomagnetic cutoff should be modified accounting for the external geomagnetic field that is perturbed
by the solar event. The proper geomagnetic cutoff used for this analysis was discussed in~\cite{GrandiICRC}.
As representative example the solar event of March 7$^{\rm th}$, 2012 is showed in Figs.~\ref{Fig::SEPFlux}--\ref{Fig::SEPRAtio}.
This was one of the most intense events
of solar cycle 24 and was correlated with two solar flares of classes X5.4 and X1.3 respectively and two fast Coronal Mass Ejections (CME). 
The time
evolution of the daily proton flux observed by AMS-02 during the March 2012 month is displayed
in Fig.~\ref{Fig::SEPRAtio} where the flux is normalized to the first day of the month at different rigidity bins. From
March 1$^{\rm st}$ until March 6$^{\rm th}$ the daily flux was stable at all rigidities. At March 7$^{\rm th}$, the increase in the
lowest rigidity bin indicates the arrival of SEP.
Forbush decrease, that was measured during the following days, even in the higher rigidity bins,
had its maximum on March 9$^{\rm th}$. After this GCR minimum, the flux gradually recovered to nominal
conditions within about 20 days.

\section{Conclusion}
In this work, we reviewed the AMS-02 GCR fluxes time evolution along the first 5 years of operation and the contribution of AMS-02 detector to Solar Physics. 
The overall flux behavior consist in a first phase of decreasing and a second part of increasing that is consistent with the solar activity cycle 
that show his maximum peak during the AMS operation. 
The minimum of the flux is the same for proton, helium and positron, while electron has minimum of the flux more then 12 month later.

The study of time evolution ratio between different species shown a clear charge sign dependent effect after the solar maximum, that can be explained 
as a transition of an opposite heliospheric magnetic field polarity. 
Particle with same charge sign show the same time evolution for the same rigidity bin. 
This leads to the conclusion that GCR propagation in the heliosphere, 
for the considered period, is a rigidity dependent process.

Finally to the solar modulation
effect, shorter timescale variations of the GCR spectra were observed due to impulsive solar events. 
Since AMS-02 will be taking data on the ISS for more than a decade, these and future data will allow
deeper studies of GCR fluxes and their time evolution over an entire solar cycle, enabling a better
understanding of the solar modulation effect and of the shorter timescale solar activity.

\bigskip 
\begin{acknowledgments}
This work is supported by ASI (Agenzia Spaziale Italiana) under contract ASI-INFN I/002/13/0.
\end{acknowledgments}

\bigskip 

\begin{thebibliography}{99} 

\bibitem{2016PhRvL.117i1103A} 
Aguilar, M., Ali Cavasonza, L., Alpat, B., et al.
Physical Review Letters, Vol.117, id.091103, 2016.

\bibitem{2013PhRvL.110n1102A} 
Aguilar, M., Alberti, G., Alpat, B., et al.,
Physical Review Letters, Vol.110, id.141102, 2013.

\bibitem{2010NIMPA.613..207A} 
Alpat, B., Ambrosi, G., Azzarello, P., et al.,
Nuclear Instruments and Methods in Physics Research A, 613, 207, 2010.

\bibitem{2011Tracker}
K. Luebelsmeyer et al., Nucl. Instrum. Methods Phys.
Res., Sect. A 654, 639, 2011.

\bibitem{2002Tracker} 
M. Aguilar et al., Phys. Rep. 366, 331, 2002.

\bibitem{2006TRD}
Doetinchem, Ph.  et al., 
Nucl. Instrum. Methods
Phys. Res., Sect. A 558, 526, 2006.

\bibitem{2013ToF} 
A. Basili, V. Bindi, D. Casadei, G. Castellini, A. Contin,
A. Kounine, M. Lolli, F. Palmonari, and L. Quadrani,
Nucl. Instrum. Methods Phys. Res., Sect. A 707, 99, 2013.

\bibitem{2014NIMPA.743...22B} 
Bindi, V., Chen, G.~M., Chen, H.~S., et al.,
Nuclear Instruments and Methods in Physics Research A, 743, 22, 2014.

\bibitem{2009ACC} 
Ph. von Doetinchem, W. Karpinski, Th. Kirn, K.
L\"{u}belsmeyer, St. Schael, and M. Wlochal, Nucl. Phys.
B, Proc. Suppl. 197, 15 (2009).

\bibitem{2010Aguilar}
M. Aguilar-Benitez et al., Nucl. Instrum. Methods Phys.
Res., Sect. A 614, 237 (2010);

\bibitem{2012Rosie}
S. Rosier-Lees et al., J. Phys. Conf.
Ser. 404, 012034 (2012)

\bibitem{2015PhRvL.114q1103A} 
Aguilar, M., Aisa, D., Alpat, B., et al., 
Physical Review Letters, Vol.114, id.171103, 2015.

\bibitem{2010IGRF}
C. C. Finlay et al., Geophys. J. Int. 183, 1216 (2010).

\bibitem{1983Unfolding}
G. D’Agostini, Nucl. Inst. Methods Phys. Res., Sect. A, 362, 487 (1995); V. Blobel, Report
DESY-84-118 (1984); A. Kondor, Nucl. Inst. Methods Phys. Res., 216, 177 (1983).

\bibitem{2007Unfolding}
J. Albert et al.,Nucl. Inst. Methods Phys. Res., Sect. A 583, 494 (2007).

\bibitem{2015ICRC_Consolandi}
Consolandi, C. PoS(ICRC2015)117, 2015

\bibitem{2015PhRvL.115u1101A} 
Aguilar, M., Aisa, D., Alpat, B., et al.,
Physical Review Letters, 115, 211101, 2015.

\bibitem{2014Aguilar}
M. Aguilar et al, Phys. Rev. Lett. 113 121102 (2014)

\bibitem{2013Adloff}
C. Adloff et al, Nucl. Instr. Meth. A 714 (2013) 147-
154

\bibitem{2014PhRvL.113v1102A} 
Aguilar, M., Aisa, D., Alpat, B., et al.,
Physical Review Letters, 113, 221102, 2014


\bibitem{2014PhRvL.113l1101A} 
Accardo, L., Aguilar, M., Aisa, D., et al.
Physical Review Letters, 113, 121101, 2014

\bibitem{2002PhR...366..331A} 
Aguilar, M., Alcaraz, J., et al., Phys. Report, Vol. 366(6), pp. 331, 2002

\bibitem{2015ICRC_Bindi}
Bindi, V., PoS(ICRC2015)108, 2015

\bibitem{GrandiICRC}
Grandi, D., Bertucci, B., Boschini,  M., PoS(ICRC2015)116, 2015

\end{thebibliography}

\end{document}